\documentclass[twocolumn,showpacs,showkeys,prl,aps]{revtex4}
\usepackage{graphicx}

\newcommand{ \be }{\begin{equation}}
\newcommand{ \ee }{\end{equation}}
\newcommand{ \bea }{\begin{eqnarray}}
\newcommand{ \eea }{\end{eqnarray}}

\begin{document}

\title{Traces of Thermalization from $p_t$ Fluctuations in Nuclear Collisions}
\author{S. Gavin}    %\email{sean@physics.wayne.edu}
\affiliation{Physics and Astronomy Department, Wayne State
University, 666 W Hancock, Detroit, MI 48201}
\date{\today}

\begin{abstract}
Scattering of particles produced in high energy nuclear collisions
can wrestle the system into a state near local thermal
equilibrium. I illustrate how measurements of the centrality
dependence of the mean transverse momentum and its fluctuations
can exhibit this thermalization.
\end{abstract}

\pacs{ 25.75.Ld, 24.60.Ky, 24.60.-k} \keywords{Relativistic Heavy
Ions, Event-by-event fluctuations.}

\maketitle

Fluctuations of the net transverse momentum have recently been
measured, with the STAR, PHENIX, NA49, and CERES experiments
reporting substantial dynamic contributions
\cite{Voloshin,PHENIX,CERES,NA49}. Such fluctuations can provide
information on collision dynamics and, perhaps, the QCD phase
transition \cite{Mrowczynski,RajagopalShuryakStephanov}.
Preliminary PHENIX and STAR data in Au+Au collisions show that
$p_t$ fluctuations increase as centrality increases
\cite{Voloshin,PHENIX}. Importantly, data from these same
experiments exhibit a strikingly similar increase in the mean
transverse momentum $\langle p_t\rangle$, a quantity unaffected by
fluctuations \cite{Adler:2003cb,VanBuren}.

I ask whether the approach to local thermal equilibrium can
explain the similar centrality dependence of $\langle p_t\rangle$
and $p_t$ fluctuations. My focus is on fluctuations, to develop
the appropriate theoretical tools and experimental observables.
Dynamic fluctuations are characterized by the observable $\langle
\delta p_{t1}\delta p_{t2}\rangle$ analyzed by STAR
\cite{Voloshin}, where it is termed $\sigma^2_{\langle
p_t\rangle,\; dynam}$, and CERES \cite{CERES}. For particles of
momenta $\mathbf{p}_1$ and $\mathbf{p}_2$, one defines
\begin{equation}\label{eq:Dynamic}
    \langle \delta p_{t1}\delta p_{t2}\rangle =
    \int\! d\mathbf{p}_{1}d\mathbf{p}_{2}\,
    {{\rho_2(\mathbf{p}_{1},\mathbf{p}_{2})}\over{\langle N(N-1)\rangle}}
    \delta p_{t1} \delta p_{t2},
\end{equation}
where $\delta p_{ti} = p_{ti}-\langle p_t\rangle$, $\langle
\cdots\rangle$ is the average over events, and $d\mathbf{p}\equiv
dyd^2p_t$. This definition exploits the relation of event-by-event
fluctuations to inclusive correlation functions discussed in
\cite{PruneauGavinVoloshin}. The pair distribution is
\begin{equation}\label{eq:2body}
\rho_2(\mathbf{p}_{1},\mathbf{p}_{2}) =
dN/d\mathbf{p}_{1}d\mathbf{p}_{2},
\end{equation}
where $\int \rho_2 d\mathbf{p}_{1}d\mathbf{p}_{2}\ = \langle
N(N-1)\rangle$ for multiplicity $N$ \cite{PruneauGavinVoloshin}.
Observe that (\ref{eq:Dynamic}) depends only on the two-body
correlation function
\begin{equation}\label{eq:corr}
r(\mathbf{p}_{1},\mathbf{p}_{2}) =
\rho_2(\mathbf{p}_{1},\mathbf{p}_{2}) -
\rho_1(\mathbf{p}_1)\rho_1(\mathbf{p}_2)
\end{equation}
with $\rho_1(\mathbf{p}) = dN/d\mathbf{p}$, since the integral
over $\rho_1\rho_1$ vanishes due to the definition of $\delta
p_{t}$. Alternative fluctuation observables $\Phi_{p_t}$,
$F_{p_t}$, and $\Delta\sigma_{p_t}$ proposed in
\cite{Mrowczynski}, \cite{Voloshin}, and \cite{PHENIX} measure
many-body correlations of all orders. These quantities are roughly
equivalent
\begin{equation}\label{eq:PhiPt}
F_{p_t} \approx \Phi_{p_t}/\sigma \approx
\Delta\sigma_{p_t}/\sigma \approx N \langle\delta p_{t1}\delta
p_{t2}\rangle/2\sigma^2
\end{equation}
when dynamic fluctuations are small compared to statistical
fluctuations $\sigma^2 = \langle p_t^2\rangle - \langle
p_t\rangle^2$ \cite{VoloshinKochRitter,PHENIX}.
%An exponential $dN/dp_{t}$ has $2\sigma^2\approx \langle p_t\rangle^2$.

\begin{figure}
%\vskip -0.4 in
%
\centerline{\includegraphics[width=3.0 in]{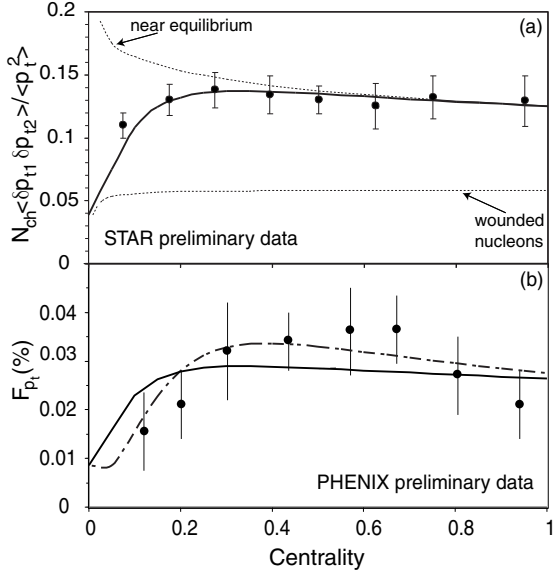}}
%
%\vskip -0.5 in
%\epsfxsize=3.50in \centerline{\epsffile{pt_fluct4.eps}}
\caption[]{(a) Dynamic $p_t$ fluctuations computed using
(\ref{eq:ptFluct}) compared to STAR data \cite{Voloshin}. (b) Same
for PHENIX data \cite{PHENIX}.} \label{fig:ptFluct}\end{figure}

STAR measurements of $p_t$ fluctuations in fig.~\ref{fig:ptFluct}a
show an increase for low multiplicities corresponding to
peripheral collisions at $s^{1/2} = 130$~GeV \cite{Voloshin}.
PHENIX measurements at 200~GeV in fig.~\ref{fig:ptFluct}b also
show such an increase for $F_{p_t}$ \cite{PHENIX}. The increase
appears to peak and possibly saturate for multiplicities
corresponding to mid-peripheral impact parameters. In addition,
the data may show a decrease for $\Delta\sigma_{p_t}$ and
$F_{p_t}$ for the most central collisions. While these
measurements are preliminary and bear large uncertainties, this
centrality dependence has already been attributed to phenomena
associated with the QCD transition \cite{strings,QCDpt}.

\begin{figure}
%\vskip -0.8 in
%
\centerline{\includegraphics[width=2.8 in]{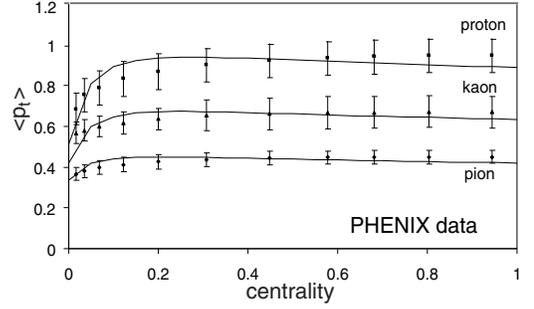}}
%
%\vskip -0.8 in
\caption[]{Average $p_t$ from (\ref{eq:meanPt}) compared to data
\cite{Adler:2003cb}. }\label{fig:PHENIX}\end{figure}

I attribute the trend in fig.~\ref{fig:ptFluct} to the onset of
thermalization in increasingly central collisions, motivated by a
similar behavior of the measured $\langle p_t\rangle$ in
fig.~\ref{fig:PHENIX} \cite{Adler:2003cb,VanBuren}. Thermalization
occurs as scattering between particles produced in the collision
drives the system toward local thermal equilibrium. The system is
characterized by a phase space density $f(\mathbf{x}, \mathbf{p},
t)$ that varies from collision event to event. As the system
approaches local equilibrium the event-averaged $\langle f\rangle$
tends toward the Boltzmann-like distribution $\langle f^e\rangle$
that varies in spacetime through the temperature $T(\mathbf{x},
t)$. I show here that thermalization alters the average transverse
momentum following
\begin{equation}\label{eq:meanPt}
    \langle p_t\rangle = \langle p_t\rangle_o S + \langle
    p_t\rangle_e (1-S),
\end{equation}
where $S$ is the probability that a particle escapes the collision
volume without scattering. Dynamic fluctuations depend on two-body
correlations and, correspondingly, are described by
\begin{equation}\label{eq:ptFluct}
    \langle \delta p_{t1}\delta p_{t2}\rangle
    =\langle \delta p_{t1}\delta p_{t2}\rangle_o S^2
    + \langle \delta p_{t1}\delta p_{t2}\rangle_e (1-S)^2.
\end{equation}
The initial quantities $\langle p_t\rangle_o$ and $\langle \delta
p_{t1}\delta p_{t2}\rangle_o$ are determined by the particle
production mechanism, while $\langle p_t\rangle_e$ and $\langle
\delta p_{t1}\delta p_{t2}\rangle_e$ depend on the state of the
system near local equilibrium.

To understand how thermalization can cause the common trends in
figs.~\ref{fig:ptFluct} and \ref{fig:PHENIX}, observe that as
centrality is increased, the system lifetime increases, eventually
to a point where local equilibrium is reached. Consequently, the
survival probability $S$ in (\ref{eq:meanPt}) and
(\ref{eq:ptFluct}) decreases from unity as the impact parameter
decreases. Both (\ref{eq:meanPt}) and (\ref{eq:ptFluct}) peak for
impact parameters near the point where equilibrium is established.
The behavior in collisions at centralities beyond that point
depends on how subsequent hydrodynamic evolution changes $\langle
p_t\rangle_e$ and $\langle \delta p_{t1}\delta p_{t2}\rangle_e$ as
the system size and lifetime increase. Systems formed in the most
central collisions can experience cooling that reduces
(\ref{eq:meanPt}) and (\ref{eq:ptFluct}).

For both the average $p_t$ and its fluctuations to \emph{increase}
during thermalization as in figs.~\ref{fig:ptFluct} and
\ref{fig:PHENIX}, both $\langle p_t\rangle_e$ and $\langle \delta
p_{t1}\delta p_{t2}\rangle_e$ must exceed the initial values. For
the average transverse momentum, this implies that the temperature
$T$ at thermalization must be quite high, since $\langle
p_t\rangle_e \propto T$. A value $\langle p_t\rangle_o\approx
350$~MeV near that measured in pp collisions implies $T\sim
400$~MeV, suggesting that partons contribute to thermalization.

In the following paragraphs I estimate $\langle \delta
p_{t1}\delta p_{t2}\rangle_o$ and $\langle \delta p_{t1}\delta
p_{t2}\rangle_e$. Next, I formulate a nonequilibrium approach
capable of treating fluctuations based on the Boltzmann-Langevin
equation in the relaxation-time approximation. Here, I sketch the
derivation of (\ref{eq:meanPt}) and (\ref{eq:ptFluct}), leaving
the details for a longer paper.

Transverse momentum and particle density fluctuations arise partly
due to the particle production mechanism, e.g., string
fragmentation. These fluctuations were measured in proton-proton
(pp) collisions \cite{ISR}. To use these pp results to estimate
$\langle \delta p_{t1}\delta p_{t2}\rangle_o$ for nuclear
collisions, I apply the wounded nucleon model to describe the soft
production that dominates $\langle p_t\rangle$ and $\langle \delta
p_{t1}\delta p_{t2}\rangle$. The charged particle multiplicity $N$
and other extensive quantities are assumed to scale linearly with
the number of participant nucleons $M$, while the intensive
one-body observable $\langle p_t\rangle$ is independent of $M$.
Centrality is determined by $N/N_{\rm max} \approx M(b)/M(0)$ for
impact parameter $b$, averaged over collision geometry.

To estimate the initial $\langle \delta p_{t1}\delta
p_{t2}\rangle$ using the wounded nucleon model, I follow the
appendix in ref.~\cite{PruneauGavinVoloshin} to obtain
\begin{equation}\label{eq:wnm}
    \langle \delta p_{t1}\delta p_{t2}\rangle_o =
    {{2\langle \delta p_{t1}\delta p_{t2}\rangle_{pp}}\over{M}}
    \left({{1+R_{pp}}\over{1+R_{AA}}}\right).
\end{equation}
The term outside the parentheses is expected because
(\ref{eq:Dynamic}) measures relative fluctuations and, therefore,
should scale as $M^{-1}$; note that pp collisions have two
participants. The term in parentheses accounts for the
normalization of (\ref{eq:Dynamic}) to $\langle N(N-1)\rangle
\equiv \langle N\rangle^2(1 + R_{AA})$ rather than $\langle
N\rangle^2$. From \cite{PruneauGavinVoloshin}, the robust variance
$R_{AA}$ satisfies
\begin{equation}\label{eq:RAA}
R_{AA} = \int\!  d\mathbf{p}_1 d\mathbf{p}_2\,
{{r(\mathbf{p}_1,\mathbf{p}_2)}\over{\langle N\rangle^2}}
= {{\langle N^2\rangle - \langle N\rangle^2 -\langle
N\rangle}\over{\langle N\rangle^2}}
\end{equation}
and scales as $R_{AA} \propto M^{-1}$. ISR measurements imply
$\langle \delta p_{t1}\delta p_{t2}\rangle_{pp}/\langle
p_t\rangle_{pp}^2 \approx 0.015$ \cite{ISR}. {\textsc{HIJING}}
gives $R_{pp}\sim 0.45$ and $R_{AA} \sim 0.0037$ for central Au+Au
for the rapidity interval $\Delta \eta = 1.5$ studied in
\cite{Voloshin}. To compare (\ref{eq:wnm}) to $N\langle \delta
p_{t1}\delta p_{t2}\rangle/\langle p_t\rangle^2$ in
fig.~\ref{fig:ptFluct}a, I assume central collisions produce
$N\approx 825$ charged particles in $\Delta\eta = 1.5$, i.e.,
$dN/d\eta \approx 550$.

Near local thermal equilibrium, dynamic fluctuations occur because
initial state fluctuations result in transient spatial
inhomogeneity that can survive thermalization. The inhomogeneity
would eventually disappear due to diffusion and viscosity, but can
be observed if freeze out is sufficiently rapid. Inhomogeneity is
essential for dynamic fluctuations, since $\langle\delta
p_{t1}\delta p_{t2}\rangle$ and $\Phi_{p_t}$ would otherwise
vanish for $\rho_2 = \rho_1\rho_1$.

To see how inhomogeneity can survive thermalization, observe that
local equilibrium is achieved when the average phase space
distribution of particles within a small fluid cell $\langle
f\rangle$ relaxes to the local equilibrium form $\langle
f^e\rangle$. The time scale for this process is the relaxation
time $\nu^{-1}$ discussed later. In contrast, density differences
\emph{between} cells must be dispersed by transport from cell to
cell. The time needed for diffusion to disperse a dense fluid mass
of size $L \sim (|\nabla n|/n)^{-1}$ is $t_{\rm d}\sim \nu
L^2/v_{\rm th}^2$, where $v_{\rm th}\sim 1$ is the thermal speed
of particles. This time can be much larger than $\nu^{-1}$ for a
sufficiently large fluid mass. The rapid expansion of the
collision system further prevents inhomogeneity from being
dispersed prior to freeze out.

Inhomogeneity produces spatial correlations: it is more likely to
find particles together near a dense fluid mass. These spatial
correlations entirely determine the phase-space correlations when
the momentum distribution at each point is thermal. I write
\begin{equation}\label{eq:ptRho1}
    r(\mathbf{p}_1,\mathbf{p}_2) = \int d\mathbf{x}_1d\mathbf{x}_2
    \mathcal{P}(\mathbf{x}_1,\mathbf{p}_1,\mathbf{x}_2,\mathbf{p}_2,t)
\end{equation}
evaluated at the freeze out proper time $\tau_F$, where the
phase-space correlation function is
\begin{equation}\label{eq:P}
\mathcal{P}_{12} \equiv \langle f_1 f_2\rangle - \langle
f_1\rangle\langle f_2\rangle - \delta_{12} \langle f_1\rangle,
\end{equation}
for $\delta_{12}= \delta(\mathbf{x}_1-\mathbf{x}_2)
\delta(\mathbf{p}_1-\mathbf{p}_2)$. A small change in density
$\delta n$ will initially drive the system from equilibrium by an
amount $\delta f^e  = f^e \delta n/n$. The corresponding phase
space correlations are described near equilibrium by
\begin{equation}\label{eq:LocalEqCorr}
\mathcal{P}_{12}^e = {{\langle f_1^e\rangle}\over{\langle
n_1\rangle}}{{\langle f_2^e\rangle}\over{\langle
n_2\rangle}}r(\mathbf{x}_1,\mathbf{x}_2),
\end{equation}
where the spatial correlation function is
\begin{equation}\label{eq:Variance}
    r(\mathbf{x}_1,\mathbf{x}_2) \equiv\langle n_1n_2\rangle - \langle
    n_1\rangle\langle n_2\rangle - \delta_{12}
    \langle n_1\rangle.
\end{equation}
The form (\ref{eq:LocalEqCorr}) ensures that both
$\mathcal{P}_{12}^e$ and $r(\mathbf{x}_1,\mathbf{x}_2)$ vanish in
global equilibrium, where particle number fluctuations obey
Poisson statistics.
I use (\ref{eq:Dynamic})  and
(\ref{eq:ptRho1})--(\ref{eq:Variance}) to find
% the induced momentum fluctuations
%
\begin{equation}\label{eq:ptRho3}
    \langle \delta p_{t1}\delta p_{t2}\rangle_e
    = \int\! d\mathbf{x}_1d\mathbf{x}_2\, r(\mathbf{x}_1,\mathbf{x}_2)
    {{\overline{\delta p_t}(\mathbf{x}_1)
    \overline{\delta p_t}(\mathbf{x}_2)}\over{\langle N(N-1)\rangle}},
\end{equation}
where the local transverse momentum excess, $\overline{\delta
p_t}(\mathbf{x}) = \int\! d\mathbf{p}\, (p_t - \langle p_t\rangle)
f(\mathbf{x},\mathbf{p})/n(\mathbf{x})$, vanishes if the collision
volume is uniform.

To estimate $\langle \delta p_{t1}\delta p_{t2}\rangle_e$ using
(\ref{eq:ptRho3}), I assume that Bjorken scaling holds and that
longitudinal and transverse degrees of freedom are independent. I
then write the transverse coordinate dependence of
(\ref{eq:Variance}) as
\begin{equation}\label{eq:corrParam}
r(\mathbf{x}_1,\mathbf{x}_2)\propto g(r_{t1})g(r_{t2})
c(|\mathbf{r}_{t1}-\mathbf{r}_{t2}|)
\end{equation}
where the density is $n(x_1)\propto g(r_{t})$. I parameterize $g$
and $c$ to be Gaussian with r.m.s.\ widths $R_t$ and $\xi$,
respectively the transverse radius and correlation length.
The momentum excess $\overline{\delta p_t}(\mathbf{x})$ in
(\ref{eq:ptRho3}) depends on the temperature profile of the
system, since $\int p_t
f(\mathbf{x},\mathbf{p})d\mathbf{p}/n(\mathbf{x})\propto T(r_t)$.
Similarly, $\langle p_t\rangle \propto \| T\|$, for the
density-weighted average $\| T\|\equiv \int
g(\mathbf{r}_t)T(\mathbf{r}_t)d\mathbf{r}_t$, so that
\begin{equation}\label{eq:momParam}
\overline{\delta p_t}(\mathbf{r}_t) = \langle p_t\rangle
[\hat{T}(r_t) -1].
\end{equation}
I parameterize $\hat{T}(r_t)= T(r_t)/\| T\|$ as Gaussian of width
$R_p$ and use $n\propto T^3$ to fix $R_p = \sqrt{3}R_t$.

The dynamic $p_t$ fluctuations near local equilibrium then satisfy
\begin{equation}\label{eq:near}
    \langle \delta p_{t1}\delta p_{t2}\rangle_{e}
    = F{{\langle p_t\rangle^2R_{AA}}\over{ 1+R_{AA}}}%F\left({{\xi_t}\over{R_t}}\right),
\end{equation}
where $R_{AA}$ is given by (\ref{eq:RAA}). The quantity $F$ is
dimensionless and depends on the ratio of the correlation length
$\xi_t$ to the transverse size $R_t$. I use (\ref{eq:corrParam})
and (\ref{eq:momParam}) to compute
\begin{equation}\label{eq:F}
F = \|
c(|\mathbf{r}_{t1}-\mathbf{r}_{t2}|)[\hat{T}(\mathbf{r}_{t1}) -
1][\hat{T}(\mathbf{r}_{t2}) - 1]\|,
\end{equation}
a double density-weighted average over $\mathbf{r}_{t1}$ and
$\mathbf{r}_{t2}$. I find $F = 0.046$ for $\xi_t/R_t = 1/6$. To
determine (\ref{eq:near}) for fig.~\ref{fig:ptFluct}a, I take
$R_{AA}=0.0037$ and $N\approx 825$ as before. I emphasize that the
{\textsc{HIJING}} $R_{AA}$ value builds in fluctuations from
resonance decay and, moreover, is roughly consistent with measured
net charge fluctuations \cite{Pruneau03}.
Let us now describe the fast local relaxation of the phase space
density $f$ to $f^e$. I start with a Boltzmann-like kinetic
equation
\begin{equation}\label{eq:Boltzmann}
    {{\partial}f/{\partial t}} +
    \mathbf{v}_\mathbf{p}\cdot\nabla f = I[f]\approx -\nu(f-f^e),
\end{equation}
approximating the collision term $I[f]$ using a single relaxation
time $\nu^{-1}$. Following \cite{Gordon,therm}, I use longitudinal
boost invariance to write the left side of (\ref{eq:Boltzmann}) as
$df/d\tau$ at fixed $p_z\tau$. Longitudinal expansion further
implies that the density satisfies $n(\tau)\propto \tau^{-1}$,
while $\langle p_t\rangle_e \propto T \propto \tau^{-\gamma}$ for
$0< \gamma<1/3$; see \cite{therm}. I then multiply both sides of
(18) by $|\mathbf{p}_t|$ and integrate over momentum to obtain
\begin{equation}\label{eq:ptLong}
\langle p_t\rangle = \langle p_t\rangle_o S +
   {{\alpha\langle p_t\rangle_e^0}\over{\alpha - \gamma}}
   \left[S^{\gamma/\alpha} -
    S\right].
\end{equation}
The survival probability is
\begin{equation}\label{eq:SurvProb}
   S = e^{-\int_{\tau_0}^{\tau_F} \nu(\tau) d\tau}
    \approx (\tau_0/\tau_F)^\alpha,
\end{equation}
where $\nu = \langle \sigma v_{\rm rel}\rangle n(\tau)$, $\alpha =
\nu_0\tau_0$ for the formation time $\tau_0$, the scattering cross
section is $\sigma$, and $v_{\rm rel}$ is the relative velocity.
For relevant values $\alpha \gg \gamma$, I approximate
(\ref{eq:ptLong}) by (\ref{eq:meanPt}) with $\langle
p_t\rangle_e\approx \langle p_t\rangle_e^0(\tau_0/\tau)^{\gamma}$.

To compute the evolution of $\langle \delta p_{t1}\delta
p_{t2}\rangle$, I obtain relaxation equations for $\mathcal{P}$.
Fluctuations due to scattering and drift are described by adding a
Langevin force to the right side of (\ref{eq:Boltzmann})
\cite{VanKampen}. On a discrete phase space lattice
$\mathbf{p}_i,\mathbf{x}_i$, the Boltzmann-Langevin equation is
\begin{equation}\label{eq:BoltzLang}
{{df_i}/{d\tau}} = -\nu (f_i-f_i^e) + \zeta_i
\end{equation}
where $\zeta_i(\tau)$ is a Langevin force. To incorporate the
effect of fluctuations near local equilibrium, I further treat
$f_i^e$ as a stochastic variable subject to an additional Langevin
force, so that $df_i^e/d\tau = \chi_i$, plus a diffusive
relaxation term that I need not specify for a diffusion time scale
$t_{\rm d} \gg \nu^{-1}$. The Langevin terms satisfy $ \langle
\zeta_i(\tau)\zeta_j(\tau^\prime)\rangle =
\nu(f+f^e)\delta_{ij}\delta(\tau-\tau^\prime)$ and $\langle
\zeta_i(\tau) \chi_j(\tau^\prime)\rangle = -\nu
f^e\delta_{ij}\delta(\tau-\tau^\prime)$, as required by detailed
balance for the relaxation-time collision term \cite{VanKampen}.
The Boltzmann equation used to compute $\langle p_t\rangle$ is the
mean value of (\ref{eq:BoltzLang}).

I use standard methods \cite{VanKampen} to obtain the following
two-body relaxation equations
\begin{eqnarray}\label{eq:relP}
  d\mathcal{P}_{ij}/d\tau &=& -2\nu \mathcal{P}_{ij} +
  \nu (\mathcal{C}_{ij}+\mathcal{C}_{ji})\\
  d\mathcal{C}_{ij}/d\tau &=& -\nu \mathcal{C}_{ij} +
  \nu \mathcal{P}_{ij}^e,
\end{eqnarray}
where I introduce the auxiliary function $\mathcal{C}_{ij} \equiv
\langle f_i f_j^e\rangle - \langle f_i\rangle\langle
f_j^e\rangle$. Observe that $\mathcal{P}_{ij} = \mathcal{C}_{ij} =
0$ in global equilibrium where the time derivatives vanish. I
solve (22) and (23) assuming that $\mathcal{C}_{ij}$ initially
vanishes and obtain $\langle \delta p_{t1}\delta p_{t2}\rangle$
from (1), (3), and (9). Equation (\ref{eq:ptFluct}) follows, but
is exact only if one neglects the time dependence of
(\ref{eq:near}) implied by $\langle p_t\rangle_e \propto
\tau^{-\gamma}$. For $\alpha \gg \gamma$, I approximate this
dependence in (\ref{eq:ptFluct}) by taking $\langle \delta
p_{t1}\delta p_{t2}\rangle_e \propto \tau^{-2\gamma}$.

I now fit this transport framework together with my earlier
assertion that near-equilibrium $p_t$ correlations are induced by
spatial inhomogeneity, i.e., eq.~(11). In the relaxation-time
approximation $\mathcal{P}^e$ is arbitrary, as is $f^e$. To deduce
either from transport theory, one must use (\ref{eq:BoltzLang})
with the full collision term $I[f]$. Following \cite{VanKampen}
yields
$\mathcal{P}_{12}^e = \langle f_1^e\rangle \langle
f_2^e\rangle\theta$, where $\theta = a_{12} +
\sum_{\mu\nu}b_{12}^{\mu\nu}p_1^\mu p_2^\nu$,
%\sum_{i,j=1}^{2}\sum_{\mu\nu}b_{ij}^{\mu\nu}p_i^\mu p_j^\nu$,
%$\mathcal{P}_{12}^e = \langle f_1^e\rangle \langle f_2^e\rangle (a
%$+ b(\mathbf{p}_1-\mathbf{p}_2)^2 + cE_1E_2)$
%
for $a$ and $b^{\mu\nu}$ functions of $x_1$ and $x_2$. In a
uniform system these coefficients are constant, so that (1), (3)
and (9) imply $\langle \delta p_{t1}\delta p_{t2}\rangle\equiv 0$,
confirming our intuition. Our physically-motivated
(\ref{eq:LocalEqCorr}) takes $\theta\approx
r(x_1,x_2)/n(x_1)n(x_2)$, which is adequate for our estimate (16).

Calculations in figs.~\ref{fig:ptFluct} and \ref{fig:PHENIX}
illustrate the common effect of thermalization on one-body and
two-body $p_t$ observables. Equation (\ref{eq:ptFluct}) together
with the computed (\ref{eq:wnm}) and (\ref{eq:ptRho3}) is in good
accord with data. The solid curves in all figures are fit to STAR
fluctuation data and $\langle p_t\rangle$ data (except for $N$, I
ignore any energy dependence). I assume $\alpha = 4$ and $\gamma =
0.15$ in central collisions, and parameterize $S(M)$ by taking
$\alpha \propto M^{1/3}$ and $\tau_F -\tau_0 \propto M^{1/2}$. In
this work it is not necessary to specify whether the equilibrating
system is partonic or hadronic. That said, in
fig.~\ref{fig:PHENIX} I take the same $\alpha$ for all species, as
appropriate for parton scattering. Measurements of $p_t$
fluctuations for identified particles can further test whether
thermalization is species independent.

In comparing to PHENIX data in fig.~\ref{fig:ptFluct}b, note that
the magnitude difference with fig.~\ref{fig:ptFluct}a follows from
the different acceptance of STAR and PHENIX. The solid curve in
fig.~\ref{fig:ptFluct}b agrees with the data within the
uncertainty, but the dashed curve shows better agreement for
$\gamma = 0.2$ and $\tau_F -\tau_0 \propto M$. While agreement
with $\langle p_t\rangle$ data for the new parameters is less
compelling than fig.~\ref{fig:PHENIX}, results still fall within
the uncertainty.

Preliminary data from refs.~\cite{Voloshin,PHENIX} and
\cite{VanBuren,Adler:2003cb} show tantalizing similarity to the
calculations. However, experimental uncertainty must be reduced to
firmly establish the low multiplicity rise as well as the behavior
at high multiplicity. Contributions to $\langle p_t\rangle$ and
$\langle \delta p_{t1}\delta p_{t2}\rangle$ not included in this
exploratory work are diffusion, collective radial flow,
Bose-Einstein (HBT) correlations, and collective hadronization.
Collective effects can be important in central collisions, where
the matter evolves after equilibration. Flow can enhance the
fluctuations, while diffusion can reduce them. HBT effects can be
experimentally estimated by cutting on each pair's relative
momentum. This contribution is of order 10\% at RHIC energy
\cite{Westfall} but may be larger at lower energies \cite{CERES}.
While resonance and hard-scattering contributions to fluctuations
are estimated by taking $R_{AA}$ in (\ref{eq:wnm}) and
(\ref{eq:near}) from \textsc{HIJING}, chemical equilibration may
modify the centrality dependence for resonance production,
altering $\langle \delta p_{t1}\delta p_{t2}\rangle$.

Experimental indications that nuclear collisions produce matter
near local equilibrium are scant and circumstantial. Any
experimental evidence of the \emph{onset} of equilibrium ---
particularly at the parton level --- will validate those
indications. Rapidity dependence measurements can distinguish the
thermalization effects proposed here from alternative explanations
\cite{QCDpt,strings}. Here, the rapidity dependence arises from
the dependence of (\ref{eq:wnm}) and (\ref{eq:near}) on $R_{AA}$,
which is itself measurable \cite{PruneauGavinVoloshin}.

\section*{Acknowledgements}

I thank C.~Pruneau, S.~Voloshin, and G.~Westfall for discussions.
This work was supported in part by the U.S. Department of Energy
under grant number DE-FG02-92ER40713.

\end{document}